\documentclass[12pt]{article}

\usepackage{amsmath}
\usepackage{amsthm} 
\usepackage{amssymb}
\usepackage{multirow}
\usepackage{color}
\renewcommand{\epsilon}{\varepsilon}

\newcommand{\Var}{\mbox{Var}}

\usepackage{amsmath}
\usepackage{amssymb}
\usepackage[ansinew]{inputenc}
\usepackage{graphicx}
\usepackage{epsfig}

\usepackage{natbib}
\newtheorem{theorem}{Theorem}[section]

\newtheorem{lemma}{Lemma}[section]

\newtheorem{remark}{Remark}[section]

\numberwithin{equation}{section}

\title{Optimal designs for comparing regression models with correlated observations}

\begin{document}

\author{
 {\small Holger Dette, Kirsten Schorning, Maria Konstantinou} \\
{\small Ruhr-Universit\"at Bochum} \\
{\small Fakult\"at f\"ur Mathematik} \\
{\small 44780 Bochum, Germany} 
}
\date{}
\maketitle

\begin{abstract}

We consider the problem of efficient statistical inference for comparing two regression curves estimated from two samples of dependent measurements. Based on a representation of the best pair of linear unbiased estimators in continuous time models as a stochastic integral, an efficient pair of linear unbiased estimators with corresponding optimal designs for finite sample size is constructed. This pair minimises the width of the confidence band for the difference between the estimated curves. We thus extend results readily available in the literature to the case of correlated observations and provide an easily implementable and efficient solution. The advantages of using such pairs of estimators with corresponding optimal designs for the comparison of regression models are illustrated via numerical examples.

\end{abstract}



\section{Introduction}

The issue of comparing two regression models that relate a common response variable to the same covariates for two different groups, respectively, arises frequently in experimental work and particularly in drug development. The conclusions drawn from such comparisons are essential for assessing the non-superiority of one model to the other or for determining whether the difference between the two models is statistically significant or not. In the latter case, the two curves describe practically the same relationship between the response variable and the covariates and the pooled sample can be used to estimate parameters of interest thus improving subsequent inference. 

Establishing the similarity of regression curves has  also been of great interest in drug dissolution studies for which several methods have been discussed in the literature [see \citet*{Yuksel}, \citet{Costa2001a}, \citet{Costa2001b} and \citet{Costa2003} among others].
 Many of the proposed methods in these references are based on measures of similarity such as  the Rescigno index or similarity factors.  
 On the more statistical side, various hypothesis tests have been proposed in the literature for assessing the equivalence of two regression curves. Linear and non-linear models with independent observations have been studied, for example, by \citet{Liu2009} and \citet*{Gsteiger2011}, respectively. Their approach is based on estimating the regression curves in the different samples and constructing confidence bands for the maximal deviation distance between these estimates. More recently, \citet{Dette-comparingcurves} propose to directly estimate the maximal deviation distance or an $L_2$-distance between the curves and to establish equivalence if the estimator is smaller than a given threshold. 

On the other hand, the efficient planning of experiments for comparing curves has not been dealt with in the literature although this would substantially improve the accuracy of the conclusions drawn regarding non-superiority or equivalence. To the best of the authors knowledge only \citet{Dette-Schorning} investigate such a design problem. They consider regression models with independent observations and search for designs that minimise the width of the simultaneous confidence band proposed by \citet*{Gsteiger2011}, for the difference of the two models. More precisely, an optimal pair of designs minimises an $L_p$-norm calculated in the common covariate region of interest, of the asymptotic variance of the difference between the two regression curves estimated via maximum likelihood. \citet{Dette-Schorning} provide explicit solutions for some commonly used dose-response models and demonstrate that using an optimal pair of designs, such as the one they propose, instead of a ``standard design'' results in the width of the confidence band to be reduced by more than $50$\%. 
 Although this improvement is impressive, the results of  \citet{Dette-Schorning} can not be  used, for example, to improve the statistical 
inference  for the  comparison of  dissolution profiles since in such studies measurements are usually taken at the same 
patient and therefore can not be considered as uncorrelated.  

The goal of the present paper is to develop efficient statistical tools (estimation and design) for the comparison of two regression curves
estimated from two samples of correlated measurements. The estimation of the parameters can easily be done by (weighted) least squares as in the uncorrelated case. However, 
it is well known that the construction of optimal designs for such estimators in the case of dependent data is a rather challenging problem 
since classical tools of convex optimisation theory are not applicable.  Solutions of exact  optimal design problem are only available for specific linear models 
[see  \citet{detkunpep2008,KisStehlik2008,harstu2010}]  and
most of the  literature is focused  on asymptotic optimal designs without
avoiding  however, the issue of non-convex optimisation [see, for example, \citet{sackylv1968}  \citet{bickherz1979},  \cite{pazmue2001},
  \cite{muepaz2003}, \citet{N1985a},
 and \citet*{zhidetpep2010} among others]. As  a consequence,  optimal designs have mainly  been determined  
  for a few one-parameter linear models. 
  
Only recently, \citet*{detpepzhi2015} and \citet*{detkonzhi2015} have made substantial progress towards overcoming the issue of non-convesx optimisation. In contrast to the prevailing design of experiments practice, according to which an optimal design is constructed for a particular estimator, these authors propose to optimise the estimator and the design simultaneously. Their strategy yields new  estimators with corresponding optimal designs which are very close to the best linear unbiased estimator, that is, the weighted least squares estimator (WLSE), with corresponding optimal design. We note that such designs require non-convex optimisation to be determined. \citet*{detkonzhi2015} improved the accuracy of the proposed class of estimators and optimal designs using a new representation of the best linear unbiased estimator in the continuous time model as a stochastic integral. They thus provide an improved solution practically non distinguishable from the WLSE (with corresponding optimal design), which is also easier to implement and applicable to a broad class of linear regression models with various covariance kernels. 

The aim of this work is to fill in the gap in the literature regarding efficient planning of experiments for comparing regression models with dependent observations. 
In Section \ref{motivation} we introduce the practical problem under investigation and provide the initial basis for the comparison of curves in the case of dependent observations. Then the framework 
for continuous-time models is set-up in Section \ref{sec2} where we obtain the best pair of linear unbiased estimators for comparing such models with dependent error processes. Finally, in Section \ref{sec3} we derive the corresponding best pair of estimators with optimal designs for finite sample size and thus answer the question of how to plan experiments for comparing curves with dependent observations. Several numerical examples are discussed in Section \ref{sec4} via which we demonstrate the benefits of our results.

\section{Comparing parametric curves}\label{motivation}

Throughout this paper we consider the practical scenario of two groups where the dependence between the response and the covariates in these groups is described by two linear regression models with dependent observations given by
\begin{equation}\label{models-finite}
Y_i(t_{i,j}) = f_i^T(t_{i,j}) \theta_i + \varepsilon_i(t_{i,j}), \quad \quad j=1, \ldots, n_i;\quad i=1,2\quad.
\end{equation}
In each group a total of $n_i$ observations are taken at the time-points $t_{i,1}, \ldots, t_{i, n_i}$ varying in a compact interval $[a,b]$ and $\{\varepsilon_i(t)| t \in [a, b]\} $ are two independent Gaussian processes with $\mathbb{E}[\varepsilon_i(t_{i,j})] =0$ and covariance kernels $K_i(t_{i,k}, t_{i,l}) = \mathbb{E}[\varepsilon_i(t_{i,k})\varepsilon_i(t_{i,l})]$ denoting the covariance between observations at the points $t_{i,k}$ and $t_{i, l}$ ($i=1, 2$, $k, l= 1, \ldots, n_i$). 
The vectors of unknown parameters $\theta_1$ and $\theta_2$ are assumed to be $m_1$- and $m_2$-dimensional respectively and the corresponding $f_i(t) = \left( f_{i,1}(t), \ldots, f_{i,m_i}(t) \right)^T$, $i=1, 2$, are vectors of continuously differentiable  linearly independent regression functions.

Let $\hat{\theta} = ( \hat{\theta}_1^T, \hat{\theta}_2^T )^T$ be a pair of linear unbiased estimators for each of the two models.  Then each estimator $\hat{\theta}_i$ is normally distributed with $\mathbb{E}[\hat{\theta}_i]=\theta_i$ and covariance matrix $\mbox{Var}(\hat \theta_i) = \Sigma_i$ (i=1, 2). 
Moreover, the prediction for the difference of the time-point $t$ satisfies
\begin{equation*} 
 f_1^T(t) \hat{\theta}_1 - f_2^T(t) \hat{\theta}_2 - (f_1^T(t) \theta_1 - f_2^T(t) \theta_2) \sim  \mathcal{N} \bigl(0, g(t,\hat{\theta}) \bigr),
\end{equation*}
where the function $g$ is defined by
\begin{align} \label{varofdif}
 g(t,\hat{\theta}) = \Var(f^T_1(t) \hat\theta_1- f^T_2(t) \hat\theta_2) &= f^T_1(t) \Var(\hat\theta_1) f_1(t) + f^T_2(t) \Var(\hat\theta_2) f_2(t) \nonumber \\
&= f_1^T(t) \Sigma_1 f_1(t)+ f_2^T(t) \Sigma_2 f_2(t).
\end{align}
We use this result and the results of \citet*{Gsteiger2011} to obtain a simultaneous confidence band for the difference of the two curves. More precisely, if $[a, b]$ is the range where the two curves should be compared, the confidence band is defined by
\begin{equation}\label{confset}
 \hat T \equiv \sup_{t \in [a,b]} \ \frac {| f_1^T(t) \hat{\theta}_1 - f_2^T(t) \hat\theta_2 - ( f_1^T(t) {\theta}_1 - f_2^T(t) \theta_2 )|}{\{  f_1^T(t) \Sigma_1 f_1(t)+ f_2^T(t) \Sigma_2 f_2(t)\}^{1/2} } \leq D,
\end{equation}
where the constant $D$ is chosen, such that $\mathbb{P}(\hat T \leq D) \approx 1-\alpha$. Note that \citet*{Gsteiger2011} propose the parametric bootstrap for this purpose.
Consequently, ``good'' estimators with corresponding good time-points $t_{1, 1}, \ldots, t_{1, n_1}, t_{2, 1}, \ldots, t_{2, n_2}$ should make  the width of this band as small as possible at each $t \in [a,b ]$. This corresponds to a simultaneous minimization of the variance in \eqref{varofdif} with respect to the choice of the linear unbiased estimators and the time-points. As pointed out by \citet{Dette-Schorning} this is only possible in rare circumstances and thus they propose to minimize an $L_p$-norm of the function $g(t,\hat{\theta})$ as a design criterion, that is
\begin{equation}\label{lp}
\| g(\cdot, \hat{\theta}) \|_p := \Big( \int_a^b |g(t,\hat{\theta})| ^p \Big)^{1/p}, \quad 1 \leq p \leq \infty, 
\end{equation} 
where the case $p=\infty$ corresponds to the maximal deviation $\| g(\cdot, \hat{\theta}) \|_{\infty} = \sup_{t \in [a,b]} |g(t,\hat{\theta})|$.

\section{Comparison in continuous time models}\label{sec2}

\citet*{detpepzhi2015} showed that the optimal design problem for most of the commonly used correlation structures can be reduced to a design problem in the regression models where the correlation structure is defined by a Brownian motion [see also Remark \ref{remarks-cont} below]. Therefore, we first examine linear regression models in their continuous time version, that is, 
\begin{equation}\label{models-cont}
Y_i(t) = f_i^T(t) \theta_i + \varepsilon_i(t), \quad \quad t\in[a,b];\quad i=1,2\quad,
\end{equation}
where $0<a<b$ and the full trajectory of both processes $Y_i=\{Y_i(t)| t \in [a,b]\}$ can be observed. The error processes $\varepsilon_i=\{\varepsilon_i(t)| t \in [a,b]\}$ are independent Brownian motions with continuous covariance kernels given by $K_i(t, t') = t \land t'$ ($i=1,2$). 
Note that we assume the processes $Y_1$ and $Y_2$ being independent. 

Following the discussion in Section \ref{motivation}, a ``good pair'' of estimators should make the $L_p$-norm of the variance given in \eqref{varofdif} as small as possible. We therefore find the best pair of estimators by minimising 
\begin{equation}\label{criterion-cont}
\mu_p(\hat \theta)=\| \Var(f^T_1(t) \hat\theta_1- f^T_2(t) \hat\theta_2)\|_p, \quad p \in [1, \infty], 
\end{equation}
with respect to all pairs of linear unbiased estimators. Recall that for $p=\infty$, we have the criterion
$\mu_\infty(\hat\theta) = \sup_{t\in[a, b]}  \Var(f^T_1(t) \hat\theta_1- f^T_2(t) \hat\theta_2)$.

 \citet*{detkonzhi2015}  showed  that  the  best linear unbiased estimators in the individual continuous time models are given by
\begin{equation}\label{blue-cont}
\hat\theta_{i, \tiny{\mbox{BLUE}}} = C_i^{-1} \left( \int_a^b \dot f_i(t) dY_i(t) + \frac{f_i(a)}{a}Y_i(a)\right) , \quad i=1, 2 \quad,
\end{equation}
where
\begin{equation}\label{C-matrix}
C^{-1}_i = \Var ( \hat\theta_{i, \tiny{\mbox{BLUE}}} ) = \left( \int_a^b \dot f_i(t) \dot{f}_i^T(t) dt + \frac{f_i(a)f_i^T(a)}{a} \right)^{-1}, \quad i=1, 2 \quad.
\end{equation}
These  estimators have minimal variance  with respect to the Loewner ordering,  that is, 
\begin{equation*}
C_i^{-1} = \Var ( \hat\theta_{i, \tiny{\mbox{BLUE}}} ) \leq \Var (\hat{\theta}_i),\quad i=1, 2 \quad,
\end{equation*}
in the Loewner ordering, for any linear unbiased estimator $\hat{\theta}_i$ in model \eqref{models-cont} ($i=1, 2$).
Theorem \ref{theorem-cont}  below shows that the best pair of linear unbiased estimators is the pair 
$(\hat\theta_{1, \tiny{\mbox{BLUE}}} ,\hat\theta_{2, \tiny{\mbox{BLUE}}} )$ of the best linear unbiased estimators in the individual models.
For its proof we first establish the following  lemma which is given as exercise $14$ of Chapter $4.3$ in \citet{borwein2000}.
 
\begin{lemma}\label{lem:saddlepoint}
Let $g:[a,b] \times \mathcal{L} \rightarrow \mathbb{R}$ be a function on the non empty set $\mathcal{L} \subset \mathbb{R}^{m_1 + m_2}$. If the point $(\bar t, \bar \theta) \in [a, b] á\times \mathcal{L} $ is a saddlepoint, that is
\begin{equation}\label{eq:sad}
g(t, \bar \theta ) \leq g (\bar t, \bar \theta) \leq g(\bar t, \theta) \quad \mbox{for all } t\in [a, b],  \, \theta \in \mathcal{L},
\end{equation}
the following equalities hold
\begin{equation}\label{eq:minmaxeq}
\inf_{\theta \in \mathcal{L}} \sup_{t \in [a,b]} g(t, \theta) = g (\bar t, \bar \theta) = \sup_{t \in [a,b]}\inf_{\theta \in \mathcal{L}}  g(t, \theta) .
\end{equation}
\end{lemma} 

{\bf Proof of Lemma \ref{lem:saddlepoint}:} 
Note that $(\bar t, \bar \theta) \in [a, b] \times \mathcal{L}$ is a saddlepoint if and only if 
\begin{equation*}
g(\bar t, \bar \theta) = \inf_{\theta \in \mathcal{L}} g(\bar t, \theta) \quad \mbox{ and } \quad g(\bar t, \bar \theta) = \sup_{t \in [a, b]} g( t, \bar \theta).
\end{equation*}
Using this formulation we have that $g(\bar t, \bar \theta) = \sup_{t \in [a, b]}  g( t, \bar \theta) \geq \inf_{\theta \in \mathcal{L}} \sup_{t \in [a,b]} g(t, \theta)$ and also $g(\bar t, \bar \theta) = \inf_{\theta \in \mathcal{L}} g(\bar t, \theta)  \leq \sup_{t \in [a,b]}\inf_{\theta \in \mathcal{L}}  g(t, \theta)$. Hence 
\begin{equation}\label{eq:sadineq}
\inf_{\theta \in \mathcal{L}} \sup_{t \in [a,b]} g(t, \theta) \leq \sup_{t \in [a,b]}\inf_{\theta \in \mathcal{L}}  g(t, \theta).
\end{equation}
On the other hand, we have that $g(t, \theta) \leq \sup_{t \in [a,b]} g(t, \theta)$ for all $t \in [a,b]$ and $\theta \in \mathcal{L}$ and thus
\begin{equation*}
\inf_{\theta \in \mathcal{L}} g(t, \theta) \leq \inf_{\theta \in \mathcal{L}}  \sup_{t \in [a,b]} g(t, \theta) \Rightarrow
\sup_{t \in [a,b]} \inf_{\theta \in \mathcal{L}} g(t, \theta) \leq  \inf_{\theta \in \mathcal{L}}  \sup_{t \in [a,b]} g(t, \theta) .
\end{equation*}
Hence
\begin{equation}\label{eq:minmaxineq}
\inf_{\theta \in \mathcal{L}} \sup_{t \in [a,b]} g(t, \theta) \geq \sup_{t \in [a,b]}\inf_{\theta \in \mathcal{L}}  g(t, \theta),
\end{equation}
and by combining \eqref{eq:sadineq} and \eqref{eq:minmaxineq} the equality \eqref{eq:minmaxeq} follows.
\hfill $\Box$

\bigskip

\begin{theorem}\label{theorem-cont}

Let $\hat\theta_{i, \tiny{\mbox{BLUE}}}$ be the best linear unbiased estimator, defined in \eqref{blue-cont} and \eqref{C-matrix}, in the corresponding continuous time model \eqref{models-cont} for $i=1, 2$. Then for any $p \in [1, \infty]$, $\hat\theta_{ \tiny{\mbox{BLUE}}} = ( \hat\theta_{1, \tiny{\mbox{BLUE}}}^T, \hat\theta_{2, \tiny{\mbox{BLUE}}}^T )^T$ is the best pair of linear unbiased estimators minimising the $L_p$-norm \eqref{criterion-cont} of the variance of the estimate of the difference between the parametric curves in the class 
\begin{equation*}
\mathcal{L} = \bigl\{\hat\theta= (\hat\theta^T_1, \hat\theta^T_2)^T | \hat\theta^T_i \mbox{ linear unbiased estimator for model \eqref{models-cont} for }  i=1, 2 \bigl\}.
\end{equation*}

\end{theorem}

{\bf Proof of Theorem  \ref{theorem-cont}:} 
Since both $\hat\theta_{i, \tiny{\mbox{BLUE}}}$ have minimal variance with respect to the Loewner ordering, it follows that $\hat\theta_{ \tiny{\mbox{BLUE}}} = ( \hat\theta_{1, \tiny{\mbox{BLUE}}}^T, \hat\theta_{2, \tiny{\mbox{BLUE}}}^T )^T$ minimises the variance of the difference between the estimated curves. That is, for any $t \in [a,b]$
\begin{equation*}
g(t, \hat\theta_{ \tiny{\mbox{BLUE}}}) \leq g(t, \hat{\theta}), \quad \mbox{for all } \hat{\theta} \in \mathcal{L} .
\end{equation*}
When $p \in [1,\infty)$, it is straightforward to check that $\hat\theta_{ \tiny{\mbox{BLUE}}}$ minimises $\mu_p(\hat{\theta}) = \| g(t, \hat{\theta})\|_p$ using the fact that the $L_p$-norm is an increasing function. Now let $p=\infty$ and also let $\hat{t} \in \arg \sup_{t\in[a,b]} g(t, \hat\theta_{ \tiny{\mbox{BLUE}}})$. It follows from the definition of $\hat{t}$ that $g(t, \hat\theta_{ \tiny{\mbox{BLUE}}}) \leq g(\hat{t}, \hat\theta_{ \tiny{\mbox{BLUE}}})$, for all $t \in [a,b]$. Therefore,
\begin{equation*}
g(t, \hat\theta_{ \tiny{\mbox{BLUE}}}) \leq g(\hat{t}, \hat\theta_{ \tiny{\mbox{BLUE}}}) \leq g(\hat{t}, \hat{\theta}), \quad \mbox{for all } t \in [a,b] \mbox{ and } \hat{\theta} \in \mathcal{L}. 
\end{equation*}
This means that $( \hat{t}, \hat{\theta}_{ \tiny{\mbox{BLUE}}} ) \in [a,b]\times \mathcal{L}$ is a saddlepoint and using Lemma \ref{lem:saddlepoint} we obtain
\begin{equation*}
\inf_{\hat{\theta} \in \mathcal{L}} \sup_{t \in [a,b]} g(t, \hat{\theta}) = g (\hat{t}, \hat{\theta}_{ \tiny{\mbox{BLUE}}} ) = \sup_{t \in [a,b]}\inf_{\hat{\theta} \in \mathcal{L}}  g(t, \hat{\theta}) .
\end{equation*}
Thus $\hat{\theta}_{ \tiny{\mbox{BLUE}}} $ minimises $\mu_{\infty}(\hat{\theta})$ in the class $\mathcal{L}$ of pairs of linear unbiased estimators. 
\hfill $\Box$
\bigskip

\begin{remark}\label{remarks-cont} {\rm 
Brownian motion is a special case of the general class of triangular kernels which are of the form
\begin{equation*}
K_i(t, t') = u_i(t) v_i(t'), \quad \mbox{for } t \leq t' \quad; i=1, 2,
\end{equation*}
for each group and the simple kernel $K_B(t,t')= t \wedge t'$ corresponding to the Brownian motion is obtained by choosing $u_i(t)=t$ and $v_i(t)=1$. \citet*{detpepzhi2015} showed that a representation of the BLUE as a stochastic integral can be obtained for any continuous time model of the form \eqref{models-cont} with a general triangular kernel. This is achieved by means of a simple non-linear transformation that reduces the model with triangular covariance kernel to a different continuous time model with Brownian motion as an error process. In particular, any model of the form \eqref{models-cont} with $\varepsilon_i(t)$ having a general triangular covariance kernel is equivalent to
\begin{equation}\label{models-trans}
\tilde{Y}_i(\tilde{t}) = \tilde{f}^T_i(\tilde{t}) \theta + \tilde{\varepsilon}_i(\tilde{t}), \quad \quad \tilde{t}\in[\tilde{a},\tilde{b}];\quad i=1,2\quad,
\end{equation}
where $\tilde{\varepsilon}_i(\tilde{t}) = \varepsilon_i(t)/ v_i(t)$, $i=1,2$, are Brownian motions on the interval $[\tilde{a},\tilde{b}]$ and
\begin{equation*}
\tilde{t} = q_i(t) := \frac{u_i(t)}{v_i(t)}, \quad \tilde{f}_i(\tilde{t}) = \frac{f_i(t)}{v_i(t)}, \quad \tilde{Y}_i(\tilde{t}) = \frac{Y_i(t)}{v_i(t)}, \quad i=1, 2.
\end{equation*} 
Hence the $\hat\theta_{i, \tiny{\mbox{BLUE}}}$ for any continuous time model \eqref{models-cont} with a general triangular covariance kernel can be obtained from the $\hat\theta_{i, \tiny{\mbox{BLUE}}}$ in the corresponding model \eqref{models-trans} by the transformation $\tilde{t}=q(t)$. Therefore, although throughout the theoretical part of this paper we focus on the covariance kernel of the Brownian motion, our methodology is valid for all triangular kernels which in fact represent the majority of covariance kernels considered in the literature. Some examples of kernels other than that of the Brownian motion are given in Section \ref{sec4} where the optimal designs are found for the transformed models \eqref{models-trans} with Brownian motion as error processes and then the design points are transformed back to the original design space $[a,b]$ via $t=q^{-1}(\tilde{t})$.
}
\end{remark}

\section{Optimal inference for comparing curves}\label{sec3}

Using the results of Section \ref{sec2} for continuous time models we can now take on our initial problem of comparing two regression curves estimated from two samples of dependent measurements, which are defined in \eqref{models-finite}. 
Following the discussion in Remark \ref{remarks-cont}, for the correlation structure in each group we assume that Cov$(Y_i(t_{i,j}), Y_i(t_{i,k})) = t_{i,j} \wedge t_{i,k}$, which corresponds to the case of a Brownian motion. Let $n = n_1 + n_2$ denote the total sample size and  define $\mathbf{Y}_i = (Y_i(t_{i,1}), \ldots, Y_i(t_{i,n_i}))^T$ 
as the vector of observations in group $i$. The corresponding weighted least squares estimator of $\theta_i$ is defined as
\begin{equation*}
\hat\theta_{i, \tiny{\mbox{WLSE}}}  = (\mathbf{X}_i^T\mathbf{\Sigma}_i^{-1}\mathbf{X}_i)^{-1}\mathbf{X}_i^T\mathbf{\Sigma}_i^{-1} \mathbf{Y}_i, \quad i=1,2 ,
\end{equation*}
where $\mathbf{X}_i=(f_{i,\ell}(t_{i,j}))^{\ell=1,\ldots,m_i}_{j=1,\ldots,n_i}$ is the $n_i\!\times\! m_i$ design matrix and
 $\mathbf{\Sigma}_i=(K_i(t_{i,j},t_{i,k}))_{j,k=1,\ldots,n_i}$ is the $n_i\!\times\! n_i$ variance-covariance matrix ($i=1, 2$). It is well known that $\hat\theta_{i, \tiny{\mbox{WLSE}}}$ is the BLUE in model \eqref{models-finite} for $i=1, 2$ and the corresponding minimal variance is given by
\begin{equation*}
 \Var (\hat\theta_{i, \tiny{\mbox{WLSE}}}) = (\mathbf{X}_i^T \mathbf{\Sigma}_i^{-1}\,\mathbf{X}_i)^{-1},\quad i=1,2.
\end{equation*}
As pointed out in the introduction the minimisation of a real-valued functional of this matrix is a demanding non-convex discrete optimisation problem and thus analytical and numerical results are rather scarce in the literature. 

An alternative to the weighted least squares estimator is proposed by \citet*{detkonzhi2015} who use an approximation of the stochastic integral involved in \eqref{blue-cont} and require the resulting estimator to be unbiased. Following this approach, we construct a numerical approximation of the stochastic integral in \eqref{blue-cont} and consider the estimators 
\begin{align}
\hat \theta_{i, n_i} &= C^{-1}_i \Bigl\{\sum_{j=2}^{n_i} \Omega_{i,j} \dot f_i(t_{i,j-1}) \bigl(Y_i(t_{i,j}) - Y_i(t_{i,j-1}) \bigr) + \frac{f_i(a)}{a} Y_i(a) \Bigr\} \nonumber \\
 &= C^{-1}_i \Bigl\{\sum_{j=2}^{n_i}\omega_{i,j} \bigl(Y_i(t_{i,j}) - Y_i(t_{i,j-1}) \bigr) + \frac{f_i(a)}{a} Y_i(a) \Bigr\} , \quad i=1,2 , \label {estimator-finite}
\end{align}
for each of the two regression models \eqref{models-finite}, where $C_i^{-1}$ is defined in \eqref{C-matrix} and $\Omega_{i,2}, \ldots, \Omega_{i,n_i}$ are $m_i \times m_i$ weight-matrices. Here $a=t_{i,1} < t_{i,2} < \ldots < t_{i,n_i-1} < t_{i,n_i} = b$, $i=1,2$, are $n_i$ design points on the time interval $[a,b]$ and $\omega_{i,2} = \Omega_{i,2} \dot{f}(t_{i,1}), \ldots, \omega_{i,n_i} = \Omega_{i,n_i} \dot{f}(t_{i,n_i-1})$ are the corresponding $m_i$-dimensional weight-vectors which have to be determined in an optimal way. We further condition on both estimators $\hat \theta_{i, n_i}$ being unbiased. It is shown in \citet*{detkonzhi2015} that unbiasedness is equivalent to the condition 
\begin{equation}\label{unbiasedness-condition}
M_i:= \int_a^b \dot{f}_i(t) \dot{f}_i^T(t) \,dt = \sum_{j=2}^{n_i} \omega_{i,j} \left( f_i^T(t_{i,j}) - f_i^T(t_{i,j-1}) \right), i=1, 2,  
\end{equation}
under which $\mbox{Var}(\hat{\theta}_{i,n_i}) = E[(\hat{\theta}_{i,n_i} - \hat{\theta}_{i,\tiny{\mbox{BLUE}}}) (\hat{\theta}_{i,n_i} - \hat{\theta}_{i,\tiny{\mbox{BLUE}}})^T] + C_i^{-1}$ [see Theorem 3.1 in \citet*{detkonzhi2015}].

For a pair   $\hat{\theta}_n = ( \hat \theta_{1, n_1} ^T , \hat \theta_{2, n_2} ^T )^T$  of linear unbiased estimators 
of the form \eqref{estimator-finite} the variance of the estimator  $ f_1^T(t) \hat \theta_{1, n_1}  - f_2^T(t) \hat \theta_{2, n_2} $
for the difference of the curves $ f_1^T(t) \theta_1 - f_2^T(t) \theta_2  $  at  a specified time-point $t \in [a,b]$ is given by
\begin{align} \label{varofdif-finite-omegas}
g_n(t,\hat{\theta}_n) &= \mbox{Var}  \big(  f_1^T(t) \hat \theta_{1, n_1}  - f_2^T(t) \hat \theta_{2, n_2} \big) = \sum_{i=1}^{2} f^T_i(t) \mbox{Var}(\hat{\theta}_{i,n_i}) f_i(t)  \nonumber \\ 
&= \nonumber 
\sum_{i=1}^{2} f^T_i(t) \Bigl\{ C_i^{-1}  E \Bigl[ \sum_{j=1}^{n_i} \int_{t_{i,j-1}}^{t_{i,j}} [\omega_{i,j}-\dot{f}(s)] \,d Y_s \sum_{k=1}^{n_i} \int_{t_{i,k-1}}^{t_{i,k}} [\omega_{i,k}-\dot{f}(s)]^T \,d Y_s \Bigl] C_i^{-1} + C^{-1}_i \Bigl\}f_i(t) \\
&= \nonumber
\sum_{i=1}^{2} f^T_i(t) \Bigl\{ C_i^{-1} \sum_{j=1}^{n_i} \int_{t_{i,j-1}}^{t_{i,j}} [\omega_{i,j}-\dot{f}(s)] [\omega_{i,j}-\dot{f}(s)]^T \,ds  C_i^{-1} + C^{-1}_i \Bigl\}f_i(t)  \\
&=    
\sum_{i=1}^{2} f^T_i(t) \Bigl\{ -C_i^{-1}M_iC_i^{-1} + \sum_{j=2}^{n_i}(t_{i,j}-t_{i,j-1}) C^{-1}_i \omega_{i,j} \omega_{i,j}^T C_i^{-1} + C^{-1}_i\Bigl\}f_i(t), 
\end{align}
where we use Ito's formula for the fourth equality. We thus need to choose the $n-4$  design points $t_{1,2}, \ldots, t_{1,n_1-1}, t_{2,2}, \ldots, t_{2,n_2-1}$ and the $n-2$ weight-vectors $\omega_{1,2}, \ldots, \omega_{1,n_1}, \omega_{2,2}, \ldots, \omega_{2,n_2}$ such that the $L_p$-norm of \eqref{varofdif-finite-omegas} is minimised in the class 
\begin{equation}\label{class-finite}
\mathcal{L}_n = \bigl\{\hat{\theta}_n= (\hat{\theta}_{1,n_1}^T, \hat{\theta}_{2,n_2}^T)^T \big| \hat{\theta}_{i,n_i} \mbox{ is of the form \eqref{estimator-finite} and satisfies condition \eqref{unbiasedness-condition} }  i=1, 2 \bigl\}.
\end{equation}
of all pairs of linear unbiased estimators of the form \eqref{estimator-finite}.
Now let $\gamma_{i,j} = \omega_{i,j} \sqrt{t_{i,j} - t_{i,j-1}}$, $j=1,\ldots,n_i; i=1,2$ and $\Gamma=(\Gamma_1^T, \Gamma_2^T)^T = (\gamma_{1,2}^T, \ldots, \gamma_{1,n_1}^T, \gamma_{2,2}^T, \ldots, \gamma_{2,n_2}^T) \in \mathbb{R}^{m_1(n_1 -1) + m_2 (n_2-1)}$. Then the variance $g_n(t,\hat{\theta}_n)$ can be rewritten as 
\begin{equation}\label{varofdif-finite-gammas}
 \varphi_n(t, \Gamma) = \sum_{i=1}^{2} f^T_i(t) \Bigl\{\tfrac{1}{a} C^{-1}_i f_i(a) f^T_i(a) C^{-1}_i + \sum_{j=2}^{n_i} C^{-1}_i \gamma_{i,j} \gamma_{i,j}^T C^{-1}_i \Bigr\} f_i(t)   .
\end{equation}
Therefore, our aim is to find an optimal pair $\Gamma^*$ such that the $L_p$-norm of \eqref{varofdif-finite-gammas} with respect to the Lebesque measure on the interval $[a,b]$, given by
\begin{equation}\label{criterion}
\mu_{p,n}(\Gamma) = \begin{cases} \left( \int_{a}^b ( \varphi_n(t, \Gamma) )^p dt\right)^{1/p}, \quad &p\in[1, \infty) \\
\sup_{t\in [a, b]} \varphi_n(t, \Gamma), \quad &p = \infty
\end{cases}
,
\end{equation}
is minimised. We note that $\mu_{p, n}$ is convex in $\Gamma$ for all $p \in [1, \infty]$.

Similar to the case of continuous time models, the optimal pair $\Gamma^*$ is the pair of optimal $\Gamma_i^*$'s which minimise the variance of the corresponding approximate estimator $\hat{\theta}_{i,n}$, given in Theorem 3.2 in \citet*{detkonzhi2015} (i=1,2).

\begin{theorem}\label{main-theorem}
Assume that each of the $m_i \times m_i$ matrices
\begin{equation*}
B_i = \sum_{j=2}^{n_i} \frac{[f_i(t_{i,j}) - f_i(t_{i,j-1})][f_i(t_{i,j}) - f_i(t_{i,j-1})]^T}{t_{i,j}-t_{i,j-1}}, \quad i=1,2,
\end{equation*} 
is non-singular. Then the optimal $\Gamma^*=((\Gamma_1^*)^T, (\Gamma_2^*)^T)^T$ minimising $\mu_{p,n}(\Gamma)$, is given by the components
\begin{equation}\label{optimal-gammas}
\gamma_{i,j}^* = M_i B_i^{-1} \frac{f_i(t_{i,j}) - f_i(t_{i,j-1})}{\sqrt{t_{i,j}-t_{i,j-1}}}, \quad j=1, \ldots,n_i; i=1,2.
\end{equation}
Moreover, the pair $\hat{\theta}_n^*= (\hat{\theta}_{1,n_1}^*, \hat{\theta}_{2,n_2}^*)$ defined in \eqref{estimator-finite} with weight-vectors given by
\begin{equation}\label{optimal-omegas}
\omega_{i,j}^* = M_i B_i^{-1} \frac{f_i(t_{i,j}) - f_i(t_{i,j-1})}{t_{i,j}-t_{i,j-1}}, \quad j=1, \ldots,n_i; i=1,2,
\end{equation}
minimises the $L_p$-norm of the function $g_n(t,\hat{\theta}_n)$ defined in \eqref{varofdif-finite-omegas} in the class $\mathcal{L}_n$ defined in \eqref{class-finite}, with respect to $\omega_{1, 2} \ldots, \omega_{1, n_1}, \omega_{2, 1}, \ldots, \omega_{2, n_2}$.
\end{theorem}

{\bf Proof of Theorem \ref{main-theorem}:}
Using similar arguments as in the proof of Theorem 3.2 in \citet*{detkonzhi2015} it can be shown that for any $v \in \mathbb{R}^{m_i}\setminus \{0\}$, $i=1,2$, each of the $\Gamma_i^*$'s, with components defined in \eqref{optimal-gammas}, minimises the function
\begin{equation*}
 v^T \Var(\hat{\theta}_{i,n}) v = v^T \Bigl\{ -C^{-1}_i M_i C^{-1}_i + C^{-1}_i \sum_{j=2}^{n_i}\gamma_{i,j} \gamma_{i,j}^T C^{-1}_i \Bigl\} v ,
\end{equation*}
subject to the constraint of unbiasedness given in \eqref{unbiasedness-condition}. Therefore, the pair $\Gamma^*=((\Gamma_1^*)^T, (\Gamma_2^*)^T)^T$ is Loewner optimal and it remains to show that $\Gamma^*$ minimises the $L_p$-norm $\mu_{p,n}(\Gamma)$ for any $p \in [1,\infty]$. The proof of this statement follows along the same lines of the proof of Theorem \ref{theorem-cont} and it is therefore omitted.
\hfill $\Box$

\bigskip

Inserting the optimal weights $\gamma_{i,j}^*$ in the function given in \eqref{varofdif-finite-gammas} and using one of the functionals in \eqref{criterion} gives an optimality criterion which is finally minimised with respect to the choice of the design points $a=t_{i,1} < t_{i,2} < \ldots < t_{i,n_i}=b$. For example, if $p=\infty$ the resulting criterion is given by
\begin{equation}\label{supcrit}
\Phi_{\infty} \Big(\{ t_{i,j} | j=1, \ldots, n_i; i=1,2 \}\Big) = \sup_{t \in [a,b]} \sum_{i=1}^2 f_i^T(t) C_i^{-1} \Bigl\{ \frac{1}{a} f_i(a) f_i^T(a) + M_i B_i^{-1}M_i \Bigl\} C_i^{-1} f_i(t) .
\end{equation}
Finally, the optimal points $a=t_{i,1}^* < t_{i,2}^* < \ldots < t_{i,n_i}^*=b$ (minimising \eqref{supcrit}) and the corresponding weights $\omega^*_{i,j}$ defined by \eqref{optimal-omegas} provide the optimal linear unbiased estimator of the form \eqref{estimator-finite} (with corresponding optimal design).

\section{Numerical examples}\label{sec4}

In this section we illustrate our methodology via several model and triangular covariance kernel examples. In particular, we 
consider the two regression curves 
\begin{align}
&Y_1 = Y_1(t) = f^T_1(t)\theta_1 + \varepsilon_1(t) =(\sin t, \cos t)^T \theta_1 + \varepsilon_1(t) \label{model1} \\
&Y_1 = Y_2(t) = f^T_2(t)\theta_2 + \varepsilon_2(t) =(\sin t, \cos t, \sin 2t, \cos 2t)^T \theta_2 + \varepsilon_2(t) \label{model2},
\end{align}
on the design space  $[a, b]= [1, 10]$, and study separately the cases of  a Brownian motion and  an  exponential 
covariance kernel of the form $K(t,t')=\exp \{ -\lambda |t-t'| \}$ for  both  error processes $ \varepsilon_1(t)$ and $ \varepsilon_2(t)$. 
 Following \citet{Dette-Schorning}, here we focus on the $\mu_{\infty}$-optimality criterion defined in \eqref{lp} since, as they 
 point out, it is probably of most practical interest and unlike the $\mu_{p}$-criteria for $p<\infty$, it is not necessarily differentiable.

\begin{table}[t]
\caption{\label{tab1} \emph{Optimal design points of five observations in each group in the interval $[1,10]$ for $\hat{\theta}_{n}^*$ when both the error processes are Brownian motions or have the same exponential covariance kernel with $\lambda=0.5$ or $\lambda=1$. The regression models $f_1$ and $f_2$ are defined in \eqref{model1} and \eqref{model2} respectively.}}
\vspace{0.5cm}
\centering
\begin{tabular}{|c|c|c|c|}
\hline
 & \multicolumn{3}{|c|}{Covariance Kernel} \\
\hline
Models & $t \wedge t'$ & $\exp \{ -0.5 |t-t'| \}$  &  $\exp \{ - |t-t'| \}$  \\
\hline
First group: $f_1$  &  [1, 3.10, 5.51, 8.40, 10] & [1, 2.87, 5.41, 8.14, 10] & [1, 2.98, 5.43, 8.03, 10]\\
\hline
Second group: $f_1$   & [1, 3.04, 5.49, 8.29, 10] &  [1, 3.15, 5.47, 8.22, 10] &  [1, 2.72, 5.48, 7.91, 10]\\
\hline
\hline
First group: $f_1$  &  [1, 3.30, 5.67, 7.34, 10] & [1, 3.20, 5.09, 8.07, 10] & [1, 2.11, 4.90, 7.77, 10]\\
\hline
 Second group: $f_2$& [1, 1.44, 5.79, 9.58, 10] &  [1, 2.43, 5.60, 5.91, 10] &  [1, 2.44, 5.29, 5.90, 10]\\
\hline
\hline
First group: $f_2$  &  [1, 1.98, 5.17, 5.51, 10] & [1, 1.54, 5.27, 9.70, 10] & [1, 5.31, 6.13, 9.00, 10] \\
\hline
Second group: $f_2$ & [1, 1.46, 5.60, 9.52, 10] &  [1, 5.15, 5.87, 9.34, 10] &  [1, 2.90, 6.63, 7.48, 10] \\
\hline
\end{tabular}
\end{table}

Throughout this section, we denote by $\hat{\theta}_{n}^* = (\hat{\theta}_{1, n_1}^*, \hat{\theta}_{2, n_2}^*)$  the best pair of linear unbiased estimators
defined by \eqref{estimator-finite}, where for  each of the combinations of models \eqref{model1} and \eqref{model2} the
optimal (vector-) weights have been found by  Theorem \ref{main-theorem} and the optimal design points  $t_{i,j}^*$ 
are determined minimising the criterion \eqref{supcrit}. For the numerical  optimization we use 
 the Particle Swarm Optimisation (PSO) algorithm [see for example \cite{Clerc2006}]
 assuming a common sample size of five observations 
in each group, that is, $n_1=n_2=5$.  Furthermore, the uniform design used in the following calculations
for a comparison  is the design which has five equally spaced design points in the interval $[1, 10]$ for each group 
 with corresponding optimal weights as in \eqref{optimal-omegas}. The $\mu_{\infty}$-optimal design points where observations should be taken in each group are given in Table \ref{tab1} for all the combinations of models and covariance kernels considered.

\begin{figure}[!htbp!]
\centering
\caption{{\it Confidence bands from the five-point $\mu_{\infty}$-optimal (solid lines) and uniform designs (dashed lines) and the true difference of the curves (dotted line). Left panel: the estimator $\hat{\theta}_{n}^*$ proposed in this paper. Right panel: the weighted least squares estimator $\hat{\theta}_{{\rm WLSE}}$. First row: model \eqref{model1} for both groups. Second row: model \eqref{model1} for first group and model \eqref{model2} for second group. Third row: model \eqref{model2} for both groups. The covariance structure is Brownian motion in all cases.}}
	\label{fig3_1} 
   \begin{minipage}{.5\textwidth}
    \centering
				\includegraphics[page=1, width=.8\textwidth]{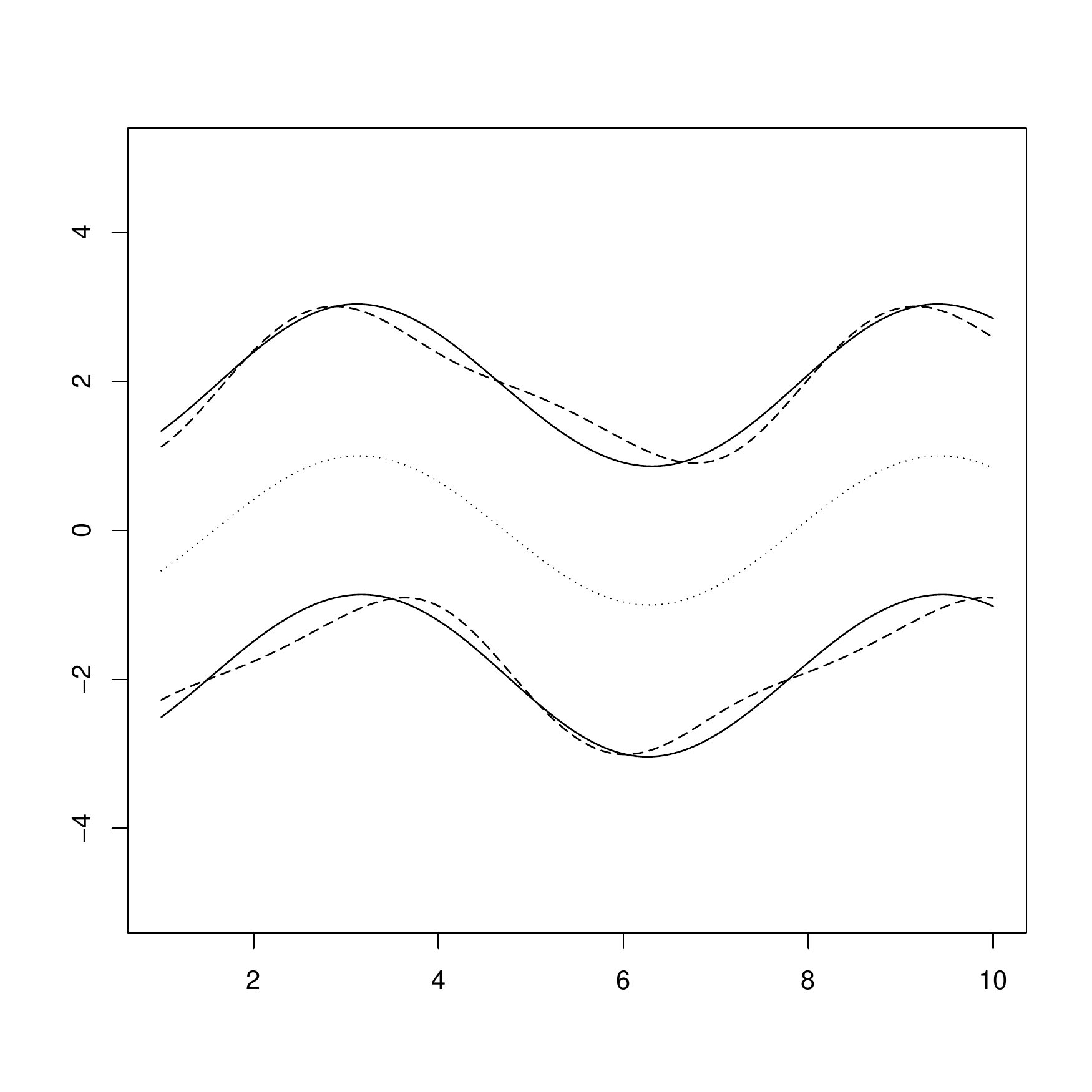} \\
		\includegraphics[page=2, width=.8\textwidth]{Approx_Conf_Brownian_110_withoutlegend} \\
		\includegraphics[page=3, width=.8\textwidth]{Approx_Conf_Brownian_110_withoutlegend} \\
		\end{minipage}%
		\begin{minipage}{.5\textwidth}
 \centering
		\includegraphics[page=1, width=.8\textwidth]{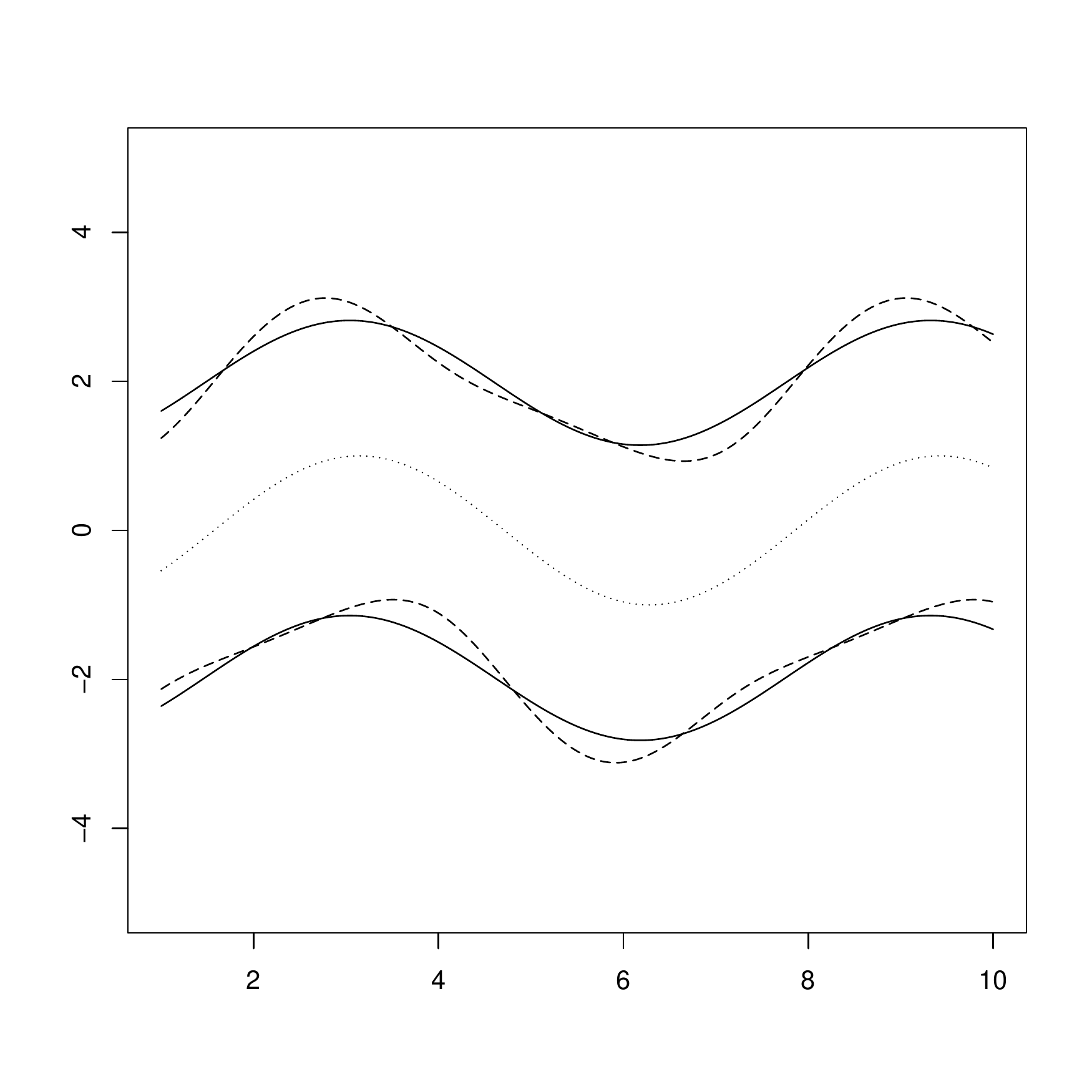} \\
		\includegraphics[page=2, width=.8\textwidth]{WLSE_Conf_Brownian_110_withoutlegend} \\
		\includegraphics[page=3, width=.8\textwidth]{WLSE_Conf_Brownian_110_withoutlegend} 
		\end{minipage}
\end{figure}

Figure \ref{fig3_1} presents the uniform confidence bands proposed by \citet*{Gsteiger2011} for the difference 
between the two regression curves. 
In the left panel  we show the results  for  the estimator proposed in this paper, while  the right panel gives the confidence 
bands obtained from  the weighted least squares estimator with a corresponding optimal design (also determined by the PSO algorithm). 
We note again that these designs
are difficult to determine because of the non-convex structure of the optimal design problem.
The depicted confidence bands are calculated as the averages of uniform confidence bands calculated by $100$ simulations. Assuming Brownian motion for both error processes, the first and last row of graphs in Figure \ref{fig3_1} 
correspond to the cases where models \eqref{model1} and \eqref{model2} respectively are used for both groups. The vectors of parameter values used for each of the groups are $(1, 1)^T$ and $(1, 2)^T$ when two models of the form \eqref{model1} are compared whereas the vectors of parameter vectors $(1, 1, 1, 1)^T$ and $(1, 2, 1, 2)^T$ were used for the last row of graphs. The middle row of graphs corresponds to the direct comparison of the two models under consideration with $\theta_{1} = (1, 1)^T$ and $\theta_{2} = (1,1, 1, 1)^T $ and assuming again Brownian motion for both error processes. In each graph, the confidence bands from the $\mu_{\infty}$-optimal or the uniform design are plotted separately using the solid and dashed lines respectively, along with the plot for the true difference $f_1^T(t)\theta_1 - f_2^T(t)\theta_2$ (dotted lines).

\begin{figure}[!htbp!]
\centering
\caption{{\it Confidence bands obtained from the estimator $\hat{\theta}_n^*$ with the five-point $\mu_{\infty}$-optimal (solid lines) and uniform designs (dashed lines) and the true difference of the curves (dotted line). Left panel: covariance kernel $\exp \{ -0.5 |t-t'| \}$. Right panel: covariance kernel $\exp \{ - |t-t'| \}$. First row: model \eqref{model1} for both groups. Second row: model \eqref{model1} for first group and model \eqref{model2} for second group. Third row: model \eqref{model2} for both groups.}}
	\label{fig3_2} 
   \begin{minipage}{.5\textwidth}
    \centering
				\includegraphics[page=1, width=.8\textwidth]{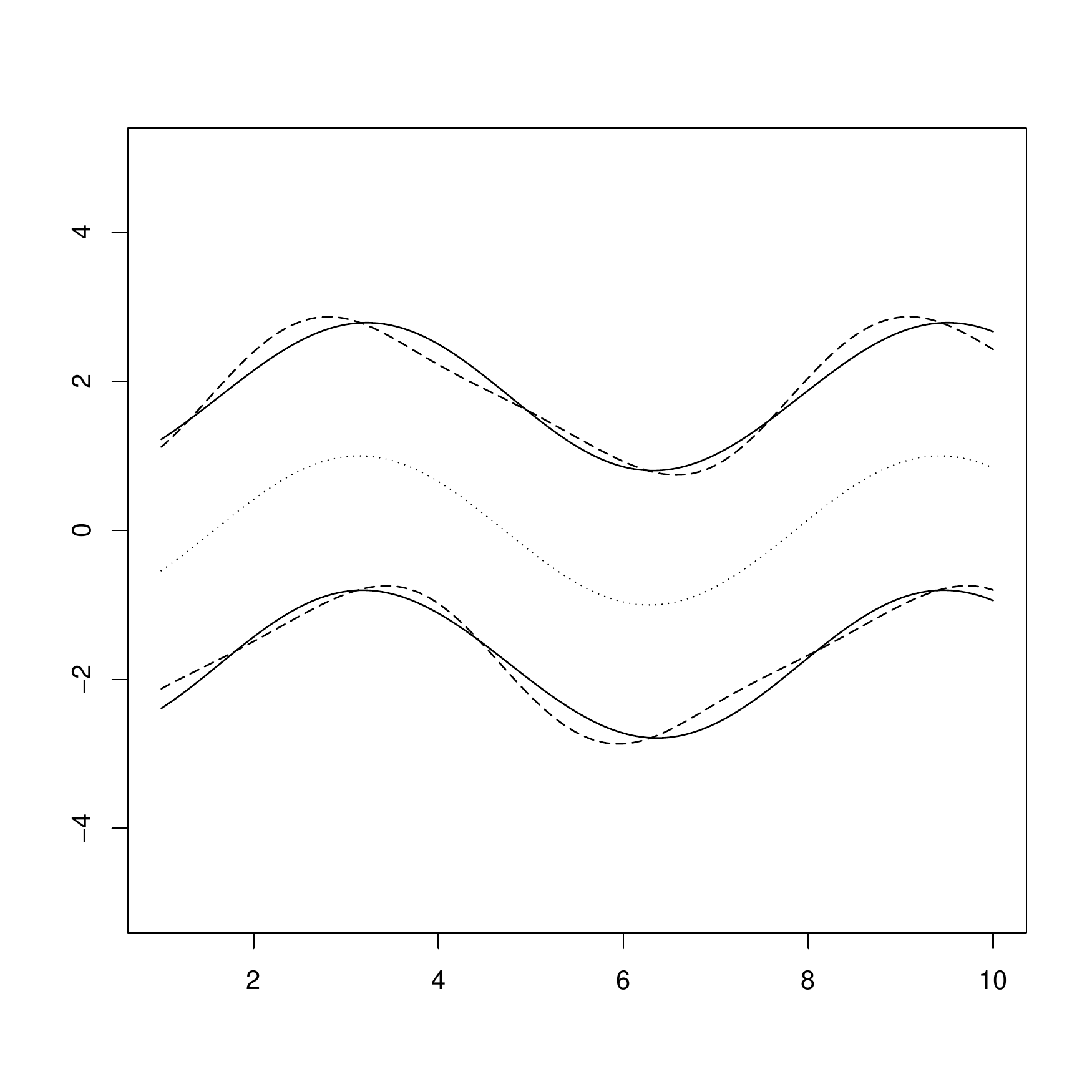} \\
		\includegraphics[page=2, width=.8\textwidth]{Approx_Conf_Exponential05_110_withoutlegend} \\
		\includegraphics[page=3, width=.8\textwidth]{Approx_Conf_Exponential05_110_withoutlegend} \\
		\end{minipage}%
		\begin{minipage}{.5\textwidth}
 \centering
		\includegraphics[page=1, width=.8\textwidth]{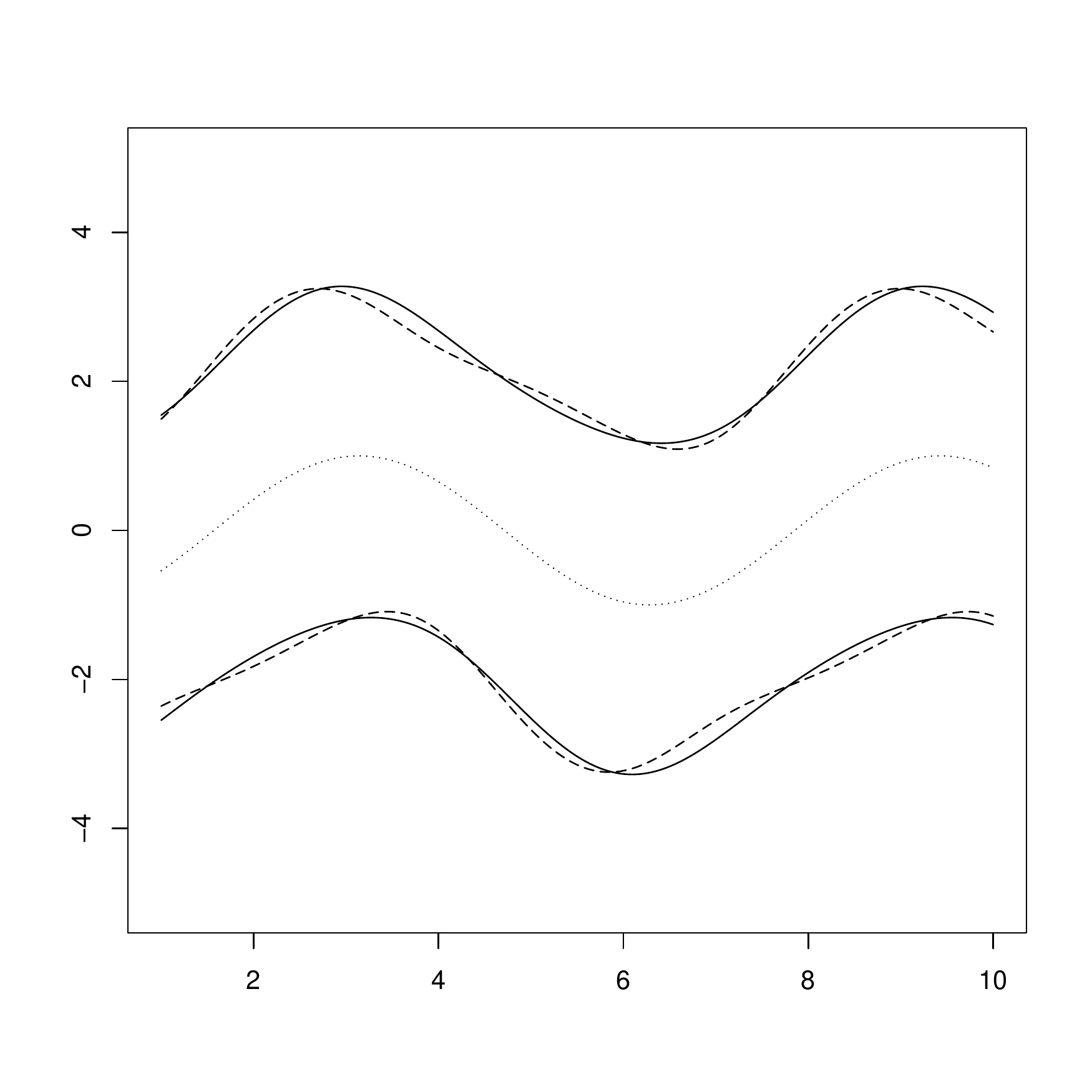} \\
		\includegraphics[page=2, width=.8\textwidth]{Approx_Conf_Exponential1_110_withoutlegend} \\
		\includegraphics[page=3, width=.8\textwidth]{Approx_Conf_Exponential1_110_withoutlegend} 
		\end{minipage}
\end{figure}

From the second and third row of graphs we observe that regardless of the estimator, if the $\mu_{\infty}$-optimal design is used instead of the uniform design the maximal width of the confidence band is reduced substantially. This is not the case for the first set of plots probably due to the small dimension of the regression function of model \eqref{model1}. We note that \citet*{detkonzhi2015} showed that for one-parameter models, equally spaced design points provide already an efficient allocation for each of the $\hat{\theta}_{i, n_i}^*$, $i=1,2$, given that the weights are chosen in an optimal way and that the derivative of the regression
is not too large. Therefore, the use of the proposed $\mu_{\infty}$-optimal design, at least up to the optimal weights, does improve inference by substantially reducing the maximum variance of the difference of the two regression curves.

By comparing the left and the right panels of Figure \ref{fig3_1} it is evident that the proposed estimator with corresponding $\mu_{\infty}$-optimal design produces similar results to those for the weighted least squares estimator with corresponding $\mu_{\infty}$-optimal design and thus in what is to follow we focus on the alternative estimator $\hat \theta_n^*$ proposed in this paper. The advantages of our methodology are also illustrated in Figure \ref{fig3_2} for the cases of the error processes of both models having the same exponential covariance kernel $K(t,t')=\exp \{ -\lambda |t-t'| \}$ with $\lambda=0.5$ (left panel) and $\lambda=1$ (right panel). As before, the maximal width of the confidence band decreases considerably when the $\mu_{\infty}$-optimal design is used at least up to the optimal weights. 

Table \ref{tab2} provides a better picture and verifies our conclusions drawn from the confidence bands plots.  Here we present the criterion values, given in \eqref{criterion}, corresponding to each of these model-kernel cases, when either the $\mu_{\infty}$-optimal or equally spaced design points are used, both with weights as in \eqref{optimal-omegas}. This gives a more direct and clearer comparison of the $\mu_{\infty}$-optimal and uniform designs the former reducing the criterion value dramatically in most of the cases.

\begin{table}[h]
\caption{\label{tab2} \emph{Criterion values $\Phi(\{ t_{i,j} | j=1, \ldots, n_i; i=1,2 \})$ for the optimal and uniform designs with five observations in each group in the interval $[1,10]$. The error processes are both Brownian motions or both have the same exponential kernel with $\lambda=0.5$ or $\lambda=1$.}}
\vspace{0.5cm}
\centering
\begin{tabular}{|c|c|c|c|c|}
\hline
\multicolumn{2}{|c|}{} & \multicolumn{3}{|c|}{Covariance Kernel} \\
\hline
  Models & Design & $t \wedge t'$ & $\exp \{ -0.5 |t-t'| \}$  &  $\exp \{ - |t-t'| \}$ \\
\hline
 First group: $f_1$   & optimal & 0.64 & 0.55 & 0.78 \\
\cline{2-5}
 Second group: $f_1$ & uniform  & 0.79 & 0.66 & 0.95    \\
\hline
\hline
First group: $f_1$  & optimal & 2.20 & 2.61 & 2.77 \\
\cline{2-5}
Second group: $f_2$  & uniform  & 27.77  & 50.95 & 68.69    \\
\hline
\hline
First group: $f_2$ & optimal & 5.06 & 1.85 & 1.83 \\
\cline{2-5}
Second group: $f_2$& uniform  & 54.91  & 25.72 &  34.68   \\
\hline
\end{tabular}
\end{table}

\medskip
\medskip

{\bf Acknowledgements.}
This work has been supported in part by the
Collaborative Research Center "Statistical modeling of nonlinear
dynamic processes" (SFB 823, Project C2) of the German Research Foundation (DFG),
the European Union Seventh Framework Programme [FP7 2007û2013] under grant agreement no 602552 (IDEAL - Integrated Design and Analysis
of small population group trials)  and the
   National Institute Of General Medical Sciences of the National
Institutes of Health under Award Number R01GM107639. The content is solely the responsibility of the authors and does not necessarily
 represent the official views of the National
Institutes of Health. 

\bigskip

\bigskip

\bibliography{comparingcurves}

\end{document}